\pdfoutput=1
\documentclass[sigconf,nonacm]{acmart}
\usepackage{booktabs}
\usepackage{tabularx}
\usepackage{array}
\usepackage[subrefformat=parens]{subcaption}
\usepackage{placeins}
\usepackage{listings}
\lstdefinestyle{tmpl}{basicstyle=\ttfamily\scriptsize,breaklines=true,
  columns=fullflexible,frame=single,framesep=4pt,xleftmargin=2pt,
  aboveskip=3pt,belowskip=0pt,keepspaces=true,escapeinside={!}{!}}
\AtBeginDocument{%
  }

\setcopyright{acmlicensed}
\copyrightyear{2026}
\acmYear{2026}
\acmConference[SIGSPATIAL '26]{the 34th ACM SIGSPATIAL International Conference on Advances in Geographic Information Systems}{November 3--6, 2026}{Riverside, CA, USA}

\begin{document}

\title{Distilling Aggregate Mobility Statistics into a Language Model
Policy for Post-Event Crowd Simulation}

\author[Tatsuya Amano, Hirozumi Yamaguchi]{%
  Tatsuya Amano$^{1,2}$, 
  Hirozumi Yamaguchi$^{1,2}$
}
\def \authors{Tatsuya Amano, Hirozumi Yamaguchi}
\affiliation{%
\institution{$^1$The University of Osaka
  \city{Suita}
\country{Japan}}
}
\affiliation{%
  \institution{$^2$RIKEN Center for Computational Science
    \city{Kobe}
  \country{Japan}}
}

\begin{abstract}
Pedestrian simulators need a behaviour rule for every agent, but privacy usually limits the data for setting one to aggregate statistics, namely zone-level device counts and origin-to-destination (OD) flows, with no individual trajectories. Such aggregates under-determine individual behaviour, because many different sets of decisions reproduce the same counts. We fine-tune a language model crowd agent so that the simulated population matches the observed destination composition, the fraction of the departing crowd heading to each point of interest. We read this target from the OD flow and reweight the model's own destination distribution onto it by iterative proportional fitting. Because fine-tuning inflates the dominant destination class, we fit the low-rank adapter to trajectories resampled to a corrected training composition that reaches the target after this inflation.
On mobile network counts from two baseball games the fine-tuned
agent runs without inference-time correction, cutting the
destination-share error by 25\%, while
the grid correlation remains similar across policies.
\end{abstract}



\maketitle

\section{Introduction}
\label{sec:intro}

When tens of thousands of people leave a large event such as a
sports match within a short time window, the way that crowd
disperses determines the safety of the surrounding area, the load
on nearby stations, and the revenue of nearby businesses. Urban
and transport planners therefore need to anticipate this dispersal
and explore how changes in facilities or transport provision would
alter the outcome \cite{10723552}. Multi-agent pedestrian
simulation~\cite{khan2024agent} can model
each person's decisions at fine spatial resolution and is realised
in widely adopted simulators such as SUMO~\cite{sumo}.

To run these simulators realistically, planners need to set each
agent's behaviour so that the simulated crowd matches what is
observed in reality. Individual trajectory data from sources such
as GPS traces or call detail records would provide direct
supervision, but privacy constraints mean that such records are
rarely released. What is typically available instead is spatially
and temporally aggregated statistics, namely device counts on a
fixed spatial grid and origin-to-destination (OD) flow matrices
between zones over time. Recovering individual behaviour from
these aggregates is ill-posed, since many different combinations
of decisions produce the same counts. Any solution must ensure
both that each agent's movements are individually plausible and
that the population as a whole remains statistically consistent
with the observed aggregates.

Existing approaches achieve statistical consistency with the
observed aggregates, for example by tuning simulator parameters
with a derivative-free
optimiser~\cite{cmaes}, by using a gravity model to match
zone-level flows~\cite{lenormand2016systematic}, or by fitting a synthetic population to known marginals with
IPF~\cite{chapuis2022generation}.
Yet they offer only a static zone-to-zone mapping, with no
model of individual behaviour that could transfer to a new
scenario. A reusable policy, a decision rule that reads an
agent's current situation and returns an action such as heading
to a station, stopping at a nearby shop, or leaving by another
exit, would let each decision depend on context such as time,
weather, and group makeup.

Recent work has turned to large language models (LLMs) as
behavioural policies in agent simulation, showing that LLM agents
reproduce believable daily routines~\cite{park2023generative,gao2024large}
and generate individually plausible mobility
patterns~\cite{llmob,jeong2025speak,LIU2026105576}.
At each decision point the simulator presents the agent's
situation as a text prompt, and the LLM reads it and returns one
executable action. Because the decision is conditioned on free-text
context, the model can respond to factors beyond what aggregate
counts capture, such as a rainy evening or a family with small
children seeking a less crowded exit. Yet the pretrained model's
destination composition, the fraction of the departing crowd
heading to each point of interest, inevitably diverges from the
observed data, leaving statistical consistency unaddressed.

In this paper we close this gap by fitting the LLM policy to the destination composition observed in the OD flow. A direct approach would train on trajectories whose class proportions match the target, but the dominant class gets amplified. A model aimed at a 74\% station share, for instance, deploys above 94\% in simulation. We avoid this overshoot by separating what the policy should produce in simulation from what it sees during training. Information projection, solved by IPF, first tilts the pretrained destination distribution onto the target while preserving the model's context-conditioned preferences within each class. We then estimate, from a few pilot training runs, how much the adaptation inflates each class, and invert this response to obtain a corrected training composition that reaches the target after inflation. A final supervised fine-tuning (SFT) pass on trajectories resampled to this corrected composition absorbs the aggregate constraint into the model weights, so the policy runs in the simulator without inference-time correction.

We evaluate on mobile network-based OD counts from two
professional baseball games at one of the largest stadiums in
Japan. After calibration to the cell counts, the grid correlation
between simulated and observed occupancy is similar across very
different behavioural priors, so this metric alone leaves them
indistinguishable. The destination composition separates them.
The fine-tuned policy cuts the destination-share error by about
a quarter relative to the untuned model, and by a further 15\%
over an inference-time correction, all in free-running simulation.

\begin{figure*}[t]
    \centering
    \includegraphics[width=0.95\linewidth]{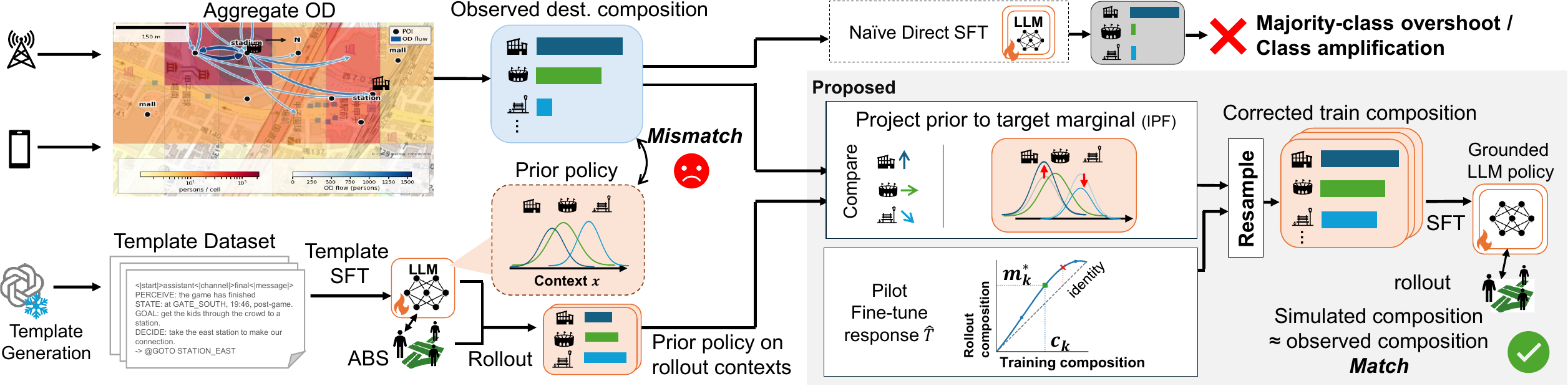}
    \caption{Proposed Method Overview.}
    \label{fig:sisa_system_overview}
\end{figure*}

\section{Overview and Problem Setting}
\label{sec:problem}

We simulate the post-event crowd as a population of agents in an
agent-based pedestrian simulator (ABS). 
We start from a base agent that follows a fixed decision
schema, fit it to the observed aggregate statistics and run the fine-tuned population to generate the post-event dispersal (Fig. \ref{fig:sisa_system_overview}).

\subsection{Agent-based simulator with an LLM policy}

Each agent is one simulated pedestrian, carrying a persona and a position in the
street network. A single language model serves as the decision policy shared by
the whole crowd, reading an agent's persona and current situation as text and
returning that agent's next action.

At each arrival or wait event, an agent reads its local situation
and responds in the OpenAI Harmony multi-channel format
(Figure~\ref{fig:template}). The \texttt{analysis} channel contains
a short, structured rationale with four labelled lines, PERCEIVE,
STATE, GOAL, and DECIDE, while the \texttt{final} channel contains
exactly one executable action. 
This design makes the model organise the relevant situation,
goal, and intended decision, and then commit to a single simulator
action. 
The simulator parses only the final
action. The rationale is discarded before the next decision, where
a fresh rationale is generated from the updated simulator state;
it therefore acts as an ephemeral reasoning scaffold rather than
part of the simulator state.

We teach this interaction schema once by supervised fine-tuning
and add $18$ dedicated action, PoI, and delimiter tokens so the
executable structure is emitted reliably.
We synthesize the format-training examples by pairing
a situation with an action sampled from the base policy for a given
persona, and a stronger model writes the surrounding rationale.
We call the resulting format-trained policy $\pi_0$. This schema
makes the behaviour readable and steerable; fitting $\pi_0$ to the
observed aggregate statistics is a separate step.

The agent and the simulator form a closed loop. We run the
population in the SUMO pedestrian simulator~\cite{sumo} over the
venue street network and advance everyone in continuous time.
Whenever an agent reaches a PoI or finishes a wait, the simulator
hands it the current situation, including any injected event. The
agent then generates a rationale and returns a final-channel action,
and the simulator executes only that action. Running the whole
population forward through this loop, generating each agent's
trajectory step by step, is a single \emph{rollout}, i.e., a
free-running simulation with no external correction. The rollout
yields the simulated behavioural-class composition that we compare
against the observed composition defined below.

\begin{figure}[t]
\centering
\begin{lstlisting}[style=tmpl]
# system: the persona, fixed
<|start|>system<|message|>Spectator at the Hanshin vs Hiroshima game,
Koshien, 2022-05-10. Retired, with daughter and two grandchildren;
unlikely to leave early, may buy a souvenir.<|end|>
# user: the situation, from the environment
<|start|>user<|message|>[19:46] The game has ended.<|end|>
# assistant: analysis channel, structured rationale
<|start|>assistant<|channel|>analysis<|message|>
PERCEIVE: the game has finished; fans are pouring out.
STATE: at !\textbf{GATE\_SOUTH}!, 19:46, post-game.
GOAL: get the kids through the crowd to a station.
DECIDE: take the east station to make our connection.
<|end|>
# assistant: final channel, action parsed by simulator
<|start|>assistant<|channel|>final<|message|>
!\textbf{@GOTO}! !\textbf{STATION\_EAST}!
<|end|>!\textbf{[STEP\_END]}!
\end{lstlisting}
\caption{One decision step in the Harmony format.  \textbf{Bold} marks the added
tokens; \texttt{\#} lines are annotations.}
\label{fig:template}
\end{figure}
Implementation details of the format pass and the aggregate-fitting
pass are given in Section~4.

\subsection{Problem setting}

The study area is a grid of $125$\,m cells and a set $\mathcal{P}$ of points of
interest (PoIs) such as train stations, shopping complexes, and venue exits. For each
ten-minute bin $t$, the operator reports a cell count
$n(c,t)$, the number of devices in cell $c$, and an
origin-to-destination (OD) flow $F(c\!\to\!c',t)$ between cells. Individual paths remain private.

We use these two statistics in two ways. The cell count $n$ says where the crowd
is at each moment; reproducing it is the standard macro check, the grid
correlation between the simulated and observed counts. The OD flow $F$ says where the crowd is
going. Aggregated over the departure window it gives the destination composition
$m^\star$, whose entry $m^\star(p)$ is the share of the departing crowd bound for PoI $p$, and this composition
is the target the policy is fitted to.

We model the crowd as $N$ agents drawn from a profile distribution
$\rho$ over age band, group makeup, and residence type. At each
step an agent with profile $z$ reads its local situation $s$ and
selects one of three actions, going to a PoI, waiting, or leaving
the area. A policy $\pi_\theta(a \mid z, s)$ generates these choices,
and we seek weights $\theta$ such that the simulated population
matches $m^\star$.

\section{Proposed Method}
\label{sec:method}

Given the format-trained $\pi_0$, we
fit it to the observed composition $m^\star$ through a second, separate
fine-tuning pass. We distinguish three compositions in this procedure. The
\emph{observed} composition $m^\star$ is read from the OD flow. The \emph{training}
composition is the class makeup of the fine-tuning set. The \emph{simulation}
composition is what the fine-tuned policy produces when run in the simulator. The
goal is to find a training composition whose simulation composition lands on
$m^\star$. The fitting has two steps. First, an information projection
computes the action distribution closest to $\pi_0$ that meets the observed composition; we call it $q^\star$, the distribution we want the policy to
produce. Second, because supervised fine-tuning shifts the simulation composition
away from the training composition, what we train on differs from
$q^\star$. We estimate this shift with a transfer map and resample the data
to a corrected training composition $\tilde{c}$ whose deployment matches $m^\star$.

We now make the two steps precise. Let $x = (z, s)$ collect the agent profile
and its local situation into a single context variable. A rollout of $\pi_0$
yields its destination distribution $\pi_0(a \mid x)$, the model's prior over
where the crowd goes before any tuning, together with the empirical set of
contexts $x$ it visits. For the information projection we hold this context
distribution fixed and neglect the shift in context occupancy that the tuning
induces in the closed loop. 
Over this distribution we solve

\begin{equation}
  \begin{aligned}
    q^\star = \arg\min_q\; &\;
    \mathbb{E}_{x}\bigl[
      D_{\mathrm{KL}}\!\bigl(q(\cdot\mid x)\,\|\,\pi_0(\cdot\mid x)\bigr)
    \bigr] \\
    \text{s.t.}\quad &\;
    \mathbb{E}_{x,\,a\sim q}[\phi_k(a)] = m^\star_k
    \quad \forall\,k,
  \end{aligned}
  \label{eq:projection}
\end{equation}

where $\phi_k(a)$ marks each action's membership in destination class
$k$ (heading to a station, to the adjacent mall, or to another exit) and $m^\star_k$ is that class's share in $m^\star$.
The solution takes the exponential-tilt form

\begin{equation}
  q^\star(a \mid x)
  = \frac{
    \pi_0(a \mid x)\,
    \exp\!\bigl(\sum_k \lambda_k\,\phi_k(a)\bigr)
  }{Z_\lambda(x)},
  \label{eq:tilt}
\end{equation}

where the multipliers $\lambda_k$ satisfy the constraints of
Eq.~\eqref{eq:projection}. This is the information
projection of $\pi_0$ onto the constraint set~\cite{csiszar1975},
found by IPF~\cite{deming1940} in milliseconds. Because the features
$\phi_k$ are class indicators, the projection acts as a class-level
offset. It preserves the support of $\pi_0$ and the relative
probabilities within each class, adjusting only the population-level
class masses to match the data.

The second step turns $q^\star$, the target we want deployed, into the
composition the model actually trains on. The two differ because fine-tuning
amplifies whichever destination class dominates the training set. A model trained
on a set composed as $q^\star$ deploys to more concentrated shares and
overshoots the target. We
therefore calibrate the resampling composition with an
empirical transfer map $T$ that sends a training composition to the
simulation composition the fine-tuned model produces. We fit
$T$ as a per-class affine relation from a few pilot fine-tunes at
differing compositions; on our data it is close to linear, for example
a station share $c$ in the data deploys to about $0.94\,c + 0.26$.
Because $T$ is per-class and monotone, we invert it in closed form to
obtain the corrected composition $\tilde{c}$ with
$T(\tilde{c}) \approx m^\star$; matching the observed $74\%$ station
share, for instance, asks for a training share near $51\%$, since
training at $74\%$ would deploy above $94\%$.

We realise $\tilde{c}$ by resampling, with no rewriting of text, keeping every
trajectory of the rarer classes and subsampling the dominant class without
replacement. A low-rank adapter trained on this set yields the fine-tuned
policy $\pi_\theta$, which brings the simulation composition toward $m^\star$ in
free-running rollout.

\section{Experimental Setup}
\label{sec:setup}

We use mobile network-based OD data for two professional baseball
games at Hanshin Koshien (10 and 11 May 2022,
official attendances $31{,}560$ and $30{,}917$).
We cross-validate over the two game days, calibrating on each
in turn and evaluating on the other, then averaging.
The observed destination composition falls into three destination
classes, a station, the adjacent commercial complex, and other exits, with mean shares $0.744$, $0.064$, and
$0.192$ across the two days.

Both SFT passes use \texttt{gpt-oss-20b} with LoRA
(rank $64$, $\alpha{=}128$), learning rate $2{\times}10^{-4}$,
effective batch size $16$, and a $4096$-token packed context on a
single NVIDIA H100 NVL GPU. The format pass trains the embeddings
of the $18$ added tokens on $1{,}756$ synthetic decision examples
whose rationales are written by \texttt{gpt-5-mini} for five epochs.
The aggregate-fitting pass resamples narrated trajectories to the
corrected composition $\tilde{c}$ and trains on $1{,}663$ examples
for three epochs.

We evaluate each policy by a free-running rollout. 
We simulate $500$ agents from a uniform distribution over personas across ten random seeds and read each agent's class from the cells its trajectory visits. An agent counts as \emph{mall} if it enters the commercial-complex cells before leaving the area, as \emph{station} if it reaches a station without entering those cells, and as \emph{other} otherwise.

All metrics are computed over the post-game window (20:00--22:10).
The grid correlation is the Pearson correlation between the
simulated and observed per-cell occupancy shares on the
$125$\,m grid ($268$ cells), where each ten-minute bin is
separately normalised to a spatial distribution and all
cell-bin pairs active in both simulation and observation
are pooled, so the metric captures the shape of the crowd's
spatial spread rather than its absolute size.
The destination-share error is the summed absolute difference
between the simulated and observed shares of the three destination
classes, where each agent is classified by the first destination
mesh its trajectory enters.

We compare \emph{Proposed} against four LLM-based variants
(\emph{no grounding}, \emph{naive SFT}, \emph{IPF at inference},
and \emph{GRPO}~\cite{grpo}),
a rule simulator tuned by CMA-ES~\cite{cmaes},
an LLM prompting agent (LLMob~\cite{llmob}),
and a classical gravity model~\cite{wilson1971}.

\section{Results}
\label{sec:results}

\begin{figure}[t]
  \centering
  \begin{subfigure}[b]{0.49\linewidth}
    \centering
    \includegraphics[width=\linewidth]{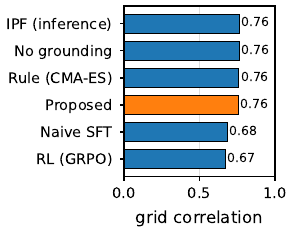}
    \caption{Grid correlation.}
    \label{fig:meshcorr}
  \end{subfigure}
  \hfill
  \begin{subfigure}[b]{0.49\linewidth}
    \centering
    \includegraphics[width=\linewidth]{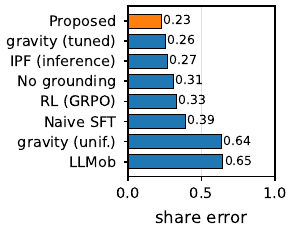}
    \caption{Destination-share error.}
    \label{fig:share_error}
  \end{subfigure}
  \caption{Grid correlation is insensitive across calibrated policies, whereas destination-share error separates them.}
  \label{fig:results}
\end{figure}

After calibration to the cell counts, the grid correlation is similar
across all behavioural priors, landing near $0.75$ for the rule, the
uniform policy, and the LLM alike (Fig.~\ref{fig:meshcorr}).
The destination composition separates them
(Fig.~\ref{fig:share_error}).
The fine-tuned policy achieves the lowest destination-share error
and is the only policy with a visible mall share, lifting
it from $0.02$ to $0.09$ against an observed $0.06$.

Training directly at the observed composition confirms the need for
the corrected training composition. Naive SFT amplifies the dominant
class and performs worse than the untuned baseline. GRPO, trained
with a cell-occupancy residual reward, also has higher destination-share
error than the fine-tuned policy and falls below the grid-correlation
band reached by the other calibrated policies.

We also tested whether the policy responds to an unseen weather prompt.
The policy is calibrated on the dry game day alone,
and the light-rain day's OD flow serves only as held-out ground truth.
We run the calibrated population on the rainy day once
with a rain sentence appended to the situation prompt
(\texttt{It has suddenly started raining.})
at the post-event onset and once without.
On the real rainy day the crowd shelters more at the commercial
complex and disperses more slowly, and conditioning the policy
on rain tracks this shift.
Late-window sheltering at the mall rises from $3$ to $59$ agents,
and the cosine similarity of the mall-occupancy curve with the
observed rainy day rises from $0.59$ to $0.74$.
A policy that samples the fixed observed composition
produces the same behaviour regardless of weather.

\section{Conclusion}
\label{sec:conclusion}

We presented a method for fitting an LLM crowd policy to
aggregate mobility statistics when individual trajectories are
unavailable. The key idea is to project the prior policy toward
the OD-derived destination composition and fine-tune on a
corrected training composition that accounts for the amplification
caused by SFT. On post-game crowd data, the calibrated policy
reduced the destination-share error and ran in free-running
simulation without inference-time correction. The results also
show that grid-count correlation alone can hide important
behavioural differences between policies.

A limitation of this work is that it targeted a post-game egress
scenario with a small set of POIs and limited context variation,
and the zero-shot weather response was demonstrated under a
single condition. As destinations and context combinations grow,
the pilot cost scales combinatorially, calling for amortised
transfer estimates or hierarchical class structures. Future work
should clarify the effective range of LLM-driven crowd simulation
across diverse venues and events, leveraging the growing
availability of open urban mobility datasets.

\begin{acks}
This work was supported by JST PRESTO Grant JPMJPR2361.
\end{acks}

\bibliographystyle{ACM-Reference-Format}
\bibliography{ref}

\end{document}